\documentclass{article}
\usepackage[T1]{fontenc} 
\usepackage[utf8x]{inputenc} 
\usepackage{ismir,amsmath,amssymb,cite,url}
\usepackage{graphicx}
\usepackage{color, caption, adjustbox}

\title{Ambisonizer: Neural Upmixing as Spherical Harmonics Generation}

\threeauthors
  {Yongyi Zang*} {Independent Researcher \\ {\tt\small zyy0116@gmail.com}}
  {Yifan Wang*\thanks{* denotes equal contribution.}} {University of Rochester \\ {\tt\small ywang419@ur.rochester.edu }}
  {Minglun Lee} {University of Rochester \\ {\tt\small minglunlee@rochester.edu  }}

\def\authorname{Y. Zang, Y. Wang and M. Lee}

\begin{document}

\maketitle
\begin{abstract}
Neural upmixing, the task of generating immersive music with an increased number of channels from fewer input channels, has been an active research area, with mono-to-stereo and stereo-to-surround upmixing treated as separate problems. In this paper, we propose a unified approach to neural upmixing by formulating it as spherical harmonics - more specifically, Ambisonic generation. We explicitly formulate mono upmixing as unconditional generation and stereo upmixing as conditional generation, where the stereo signals serve as conditions. We provide evidence that our proposed methodology, when decoded to stereo, matches a strong commercial stereo widener in subjective ratings. Overall, our work presents direct upmixing to Ambisonic format as a strong and promising approach to neural upmixing. A discussion on limitations is also provided.
\end{abstract}
\section{Introduction}\label{sec:introduction}

Channel upmixing is a technique that enables better audio playback of fewer channels audio on systems equipped with more channels. It includes two primary categories: converting mono to stereo and stereo to surround. Mono-to-stereo upmixing allows monophonic content to be enjoyed on stereo playback devices like headphones and stereo speakers, while stereo-to-surround upmixing enables stereo content to be experienced on home theater or surround sound systems.

Recent advancements in microcontroller production, head-tracking technology, and head-related transfer functions (HRTFs) have facilitated spatial content playback on consumer headphones using channel-agnostic playback formats~\cite{rumsey2012spatial, bosman2024effect}. These formats encode audio information within a spatial audio field and utilize signal processing techniques on the playback device to convert the audio into the necessary channel format. This method provides a more adaptable and versatile audio playback experience, as the audio can be tailored to the specific playback system and the listener's preferences. The trend of channel-agnostic formats is on the rise, with popular music and video streaming platforms like Apple Music and YouTube adopting this technology.

Despite the growing popularity of channel-agnostic playback, research on channel-agnostic upmixing is limited. A similar approach involves using source separation methods to isolate individual elements or tracks and position them at specific locations within the spatial audio field~\cite{lagrange2007semi, fitzgerald2011upmixing, shim2009stereo, ibrahim2018primary}. However, the quality of separation in current models is a constraint, as they often can only separate a limited number of predetermined tracks\footnote{Models such as~\cite{lin2021unified, chen2022zero} can separate undetermined number of tracks, yet their performances lack significantly behind their 
 predetermined counterparts, thereby is currently unusable for neural upmixing.}\cite{pereira2023moisesdb} and introduce noticeable artifacts after upmixing~\cite{cano2016evaluation, pons2021upsampling}. Moreover, the manual placement of elements in the spatial audio field necessitates human expertise, limiting the practicality of this approach. This limitation highlights the necessity for further research and advancement in channel-agnostic upmixing to enhance the quality and effectiveness of spatial audio playback on consumer headphones.

In response to this need, we propose \textbf{Ambisonizer}, a novel paradigm that leverages spherical harmonics to create channel-agnostic neural upmixing. By leveraging the Ambisonic format, we directly generate an Ambisonic first-order upmix from a mono sound file. Under mono-to-any upmixing scenarios, we treat the problem as unconditional generation; under stereo-to-any upmixing scenarios, we downmix the stereo signal into mono, and treat the stereo signal as a spatial condition to the generation process. To the best of our knowledge, our work proposes the first framework that allows for such mono-to-any and stereo-to-any neural upmixing. 

Through subjective evaluations, we demonstrate that both mono-to-any and stereo-to-any Ambisonizer generation results, when downmixed to stereo, match a strong commercial mono-to-stereo upmixing baseline, with the added benefit of being channel-agnostic. We also discuss the limitations of using the Ambisonic B-format as a middle format for channel-agnostic upmixing. To facilitate future research, our code, model artifacts, and data generation pipeline will be open-sourced soon at \url{https://ambisonizer.netlify.app}.




\section{The Ambisonic format}
\subsection{First-Order Ambisonics}
Ambisonics is a multichannel format designed to capture and reproduce the spatial characteristics of sound fields. The encoding of audio information into the Ambisonic format typically begins with first-order Ambisonics, which utilizes four spherical harmonic channels: $W$, $X$, $Y$, and $Z$ (known collectively as the Ambisonic B-format). The $W$ channel is also considered as the zeroth-order Ambisonics, which represents the omnidirectional component of the sound field, proportional to the acoustic pressure $p(t)$, similar to how an omnidirectional microphone captures to capture sound from all directions. $X$, $Y$ and $Z$ channels capture sound from $X$-axis, $Y$-axis and $Z$-axis correspondingly in a similar fashion as 8-figure microphones. Together, these channels encode both the magnitude and the directional information of the sound at any given point within the sound field. If elevation information is not needed during decoding, the $Z$ channel information may be omitted in first-order Ambisonics~\cite{arteaga2015introduction}.


Ambisonic decoders are used for rendering the Ambisonic format to specific speaker layouts. Accurate and efficient implementations of decoders are an active area of research~\cite{peters2020auralising}. However, it is currently hard for any Ambisonic decoders to recreate the sound field perfectly. The finite number of spherical harmonics coefficients used in Ambisonics commonly lead to truncation artifacts, affecting the accuracy of sound field recreation. As a result, many decoders prioritize either physical or perceptual accuracy. For instance, an In-Phase decoder could greatly reduce localization artifacts but may not provide the best physical accuracy compared to other methods~\cite{murillo2014evaluation}.

\subsection{Higher Order Ambisonics}
\begin{figure}[hbt!]
    \centering
    \includegraphics[width=\linewidth]{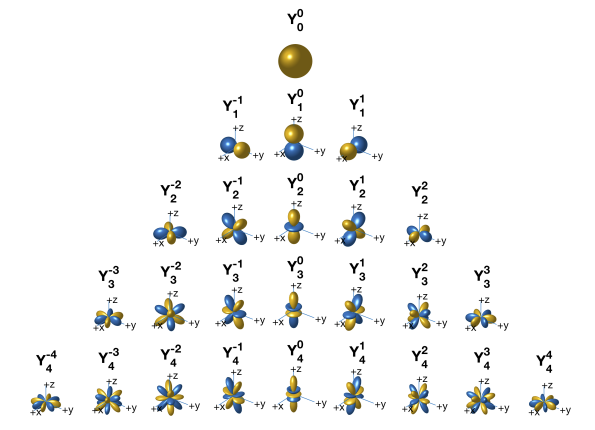}
    \caption{Higher Order Ambisonic ($4^{th}$ order)\cite{eigenbeam2023}}
    \label{fig:HOA}
\end{figure}

The first-order Ambisonic B-format represents the sound field using only 4 channels, which provides limited spatial information. This limitation places a greater burden on decoders, which are responsible for rendering the audio to the desired playback channel configuration. As a result, perceptual deficiencies, such as poor localization accuracy and coloration, may occur~\cite{mccormack2019parametric}. To mitigate these issues, Higher-Order Ambisonics (HOA)~\cite{daniel2004further} has been introduced, allowing for higher spatial resolution in Ambisonic data. In HOA, the spherical harmonics are arranged symmetrically, centering around the \( z \)-rotationally symmetric component for each order. The components to the left of the center represent sine-based horizontal components, while the components to the right represent cosine-based horizontal components. An example illustration for $4^{th}$-order Ambisonic is illustrated in Figure~\ref{fig:HOA}.


\section{The Ambisonizer Model}\label{sec:typeset_text}
The Ambisonizer model is based on the intuition that existing mono-to-stereo and stereo-to-surround upmixing models implicitly generate spherical harmonics. This is because these models require an inherent understanding of spherical harmonics to successfully upmix audio to a higher number of channels. Consequently, we propose directly upmixing the input signal to Ambisonic B-format, which enables more explicit way of generating spherical harmonics. This approach allows the utilization of existing Ambisonic decoders to render the upmixed audio to specific channel layouts, providing flexibility and compatibility with various playback systems.

\begin{figure}[hbt!]
    \centering
    \includegraphics[width=\linewidth]{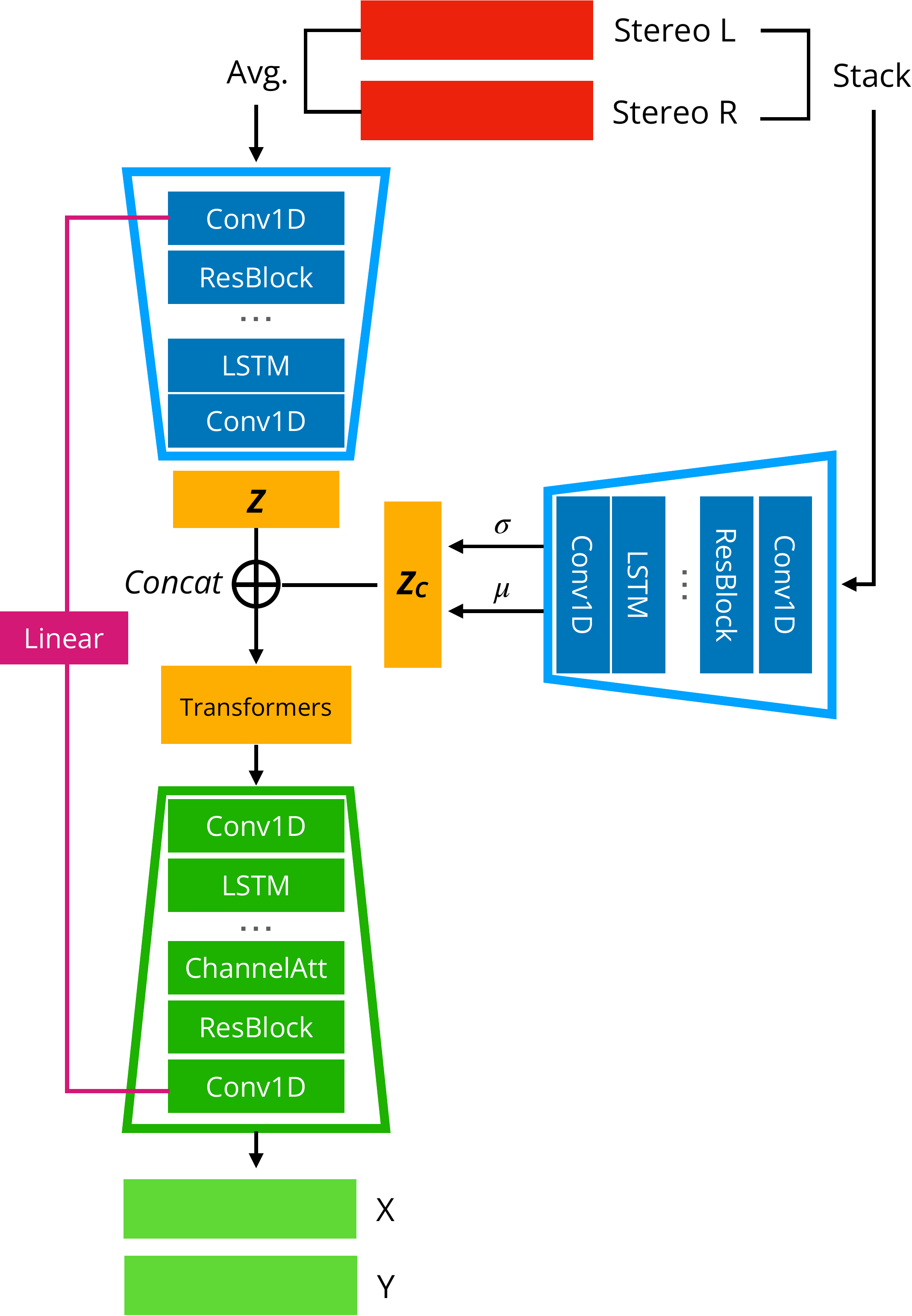}
    \caption{The Ambisonizer model architecture. Blue blocks denote encoders, and green block denotes the decoder. Best viewed in color.}
    \label{fig:model_arch}
\end{figure}

Figure~\ref{fig:model_arch} illustrates the overall framework of the proposed Ambisonizer model, with input $Y = (Y_L, Y_R)$ and output $Y' = (Y_W, Y_X, Y_Y)$. We start by averaging the two channels to obtain $Y_{mono} = \frac{1}{2} (Y_L + Y_R)$, which we posit is equal to $Y_W$. $Y_W$ serves as input to the audio encoder, and the stereo signal $Y$ is used as input to the spatial information encoder, deriving $Z$ and $Z_C$ correspondingly. The results are combined and processed using transformer encoder layers, then passed through a decoder to obtain $Y_X$ and $Y_Y$, thereby obtaining the complete first-order Ambisonics (excluding elevation information $Y_Z$). $Y_W$, $Y_X$, and $Y_Y$ directly form the Ambisonic B-format $W$, $X$, and $Y$ channels, which can be directly utilized for decoding. 

\subsection{Audio and Spatial Encoders}
The audio and spatial encoders in our model are inspired by the designs presented in~\cite{tagliasacchi2020seanet} and~\cite{defossez2022high}. Given a raw waveform as an input, the audio encoder first applies a 1D convolution with a kernel size of 7 and 32 output channels. The output then passes through a series of residual convolution blocks, each consisting of a residual-connected 1D convolution followed by a downsampling block. After each downsampling operation, the number of channels is doubled to capture more abstract features. The residual block is repeated four times with strides of $(2, 4, 5, 8)$. The final output of the audio encoder is obtained by passing the feature map through a two-layer LSTM and a 1D convolution with a kernel size of 7. 

The spatial encoder is additionally treated as a variational autoencoder (VAE)~\cite{kingma2013auto}, with its output reparameterized to follow a unit Gaussian distribution, $\mathcal{N}(0, 1)$. This reparameterization allows for direct sampling from the Gaussian distribution when stereo information is not provided, enabling the unconditional generation of Ambisonic audio from mono input. 

\subsection{Bottleneck}

The bottleneck of our model consists of a series of transformer encoder layers~\cite{vaswani2017attention}. We concatenate the latent representation $Z$ obtained from the audio encoder and the spatial embedding $Z_C$ obtained from the spatial encoder. The concatenated vector is then fed into the transformer encoder layers. We use 8 attention heads and 8 transformer encoder layers to capture complex dependencies and learn a rich representation of the audio and spatial information.

\subsection{Decoder}

The decoder of our model mirrors the structure of the encoder but with transposed 1D convolutions instead of strided convolutions. This design choice allows the decoder to gradually upsample the bottleneck representation and generate the output Ambisonic audio. 

To further enhance the decoder's ability to capture cross-channel dependencies, we introduce channel attention mechanisms~\cite{hu2018squeeze} before each residual block. Channel attention allows the model to adaptively weigh the importance of different channels at each stage of the decoding process. This is achieved by learning a set of channel-wise weights that are applied to the feature maps. The channel attention mechanism can be formulated as follows:

\begin{equation}
F_{att} = \sigma(W_2(\text{ReLU}(W_1(F_{avg}))))\odot F
\end{equation}

where $F_{att}$ is the attended feature map, $F$ is the input feature map, $F_{avg}$ is the channel-wise average pooled feature map, $W_1$ and $W_2$ are learnable linear layers, $\sigma$ is the sigmoid function, and $\odot$ is element-wise multiplication.

\subsection{Loss Function}
We use the Evidence Lower Bound (ELBO) loss to train the Ambisonizer model, where reconstruction loss is balanced with a regularization term to encourage the learned distribution of the latent variables produced by the spatial encoder to follow unit Gaussian $\mathcal{N}(0, 1)$. It is defined as:

\begin{align}
\mathcal{L}_{\text{ELBO}} = & -\mathbb{E}_{q(Z_C \mid Y)}[\log p(Y \mid Z_C)] \nonumber \\
& + \mathbb{KL}\big(q(Z_C \mid Y) \,\|\, p(Z_C)\big)
\end{align}

where $Y$ is the input stereo audio, $Z_C$ is the spatial embedding, $q(Z_C \mid Y)$ is the approximate posterior distribution learned by the encoder, $p(Y|Z_C)$ is the likelihood of the input given the spatial embedding, and $p(Z_C)$ is the prior distribution of the spatial embedding. $\mathbb{KL}(q(Z_C \mid Y) \,\|\, p(Z_C))$ is the Kullback-Leibler divergence between the approximate posterior $q(Z_C \mid Y)$ and the prior $p(Z_C)$.

For the reconstruction term, we use a combination of a multi-resolution Short-Time Fourier Transform (STFT) loss, and a mean squared error ($L_2$) loss on the waveform. 
The multi-resolution STFT loss~\cite{yamamoto2020parallel} is used to capture the time-frequency characteristics of the generated audio. It is computed by taking the STFT of the generated and target Ambisonic audio at multiple resolutions and comparing their magnitudes. The STFT loss is defined as:

\begin{equation}
\mathcal{L}_{STFT} = \sum_{r=1}^{R} \left\| |\text{STFT}_{r}(\hat{y})| - |\text{STFT}_{r}(y)| \right\|_{1}
\end{equation}

where $\hat{y}$ is the generated Ambisonic channels, $y$ is the target Ambisonic channels, $\text{STFT}_{r}$ denotes the STFT operation at resolution $r$, and $R$ is the total number of resolutions. We follow~\cite{yamamoto2020parallel} and select $R=3$, under the FFT sizes of $[512, 1024, 2048]$ with window sizes of $[240, 600, 1200]$ and hop sizes of $[50, 120, 240]$.

The $L_2$ loss on the waveform is used to ensure that the generated audio closely matches the target audio in the time domain. It is defined as:

\begin{equation}
\mathcal{L}_{2} = \left\| \hat{y} - y \right\|_{2}^{2}
\end{equation}

Using a combination of time domain and time-frequency domain loss can allow the model to capture both magnitude and phase information effectively, while not overfit to low frequencies~\cite{defossez2019music, defossez2022high}. Since phase information is crucial for Ambisonic audio, we scale the $L_2$ loss by a factor of 10 to emphasize its importance. The final loss function is defined as:

\begin{equation}
\mathcal{L} = \mathcal{L}_{STFT} + 10 \cdot \mathcal{L}_{2} + \mathbb{KL}\big(q(Z_C \mid Y) \,\|\, p(Z_C)\big)
\end{equation}

\section{Experiments}
\subsection{Synthesizing first-order Ambisonic data}
To train the Ambisonizer model, we require data pairs consisting of $Y = (Y_L, Y_R)$ and $Y' = (Y_W, Y_X, Y_Y)$. We synthesize $Y'$ directly using Ambisonic impulse response (IR) datasets. Mono sound sources are encoded with the first-order Ambisonic IR without $Z$ channel, denoted as $(IR_W, IR_X, IR_Y)$, to obtain $Y'$. For a given sound source $S$ at azimuth $\theta$, we convolve it with $IR_W$, $IR_X$ and $IR_Y$ and calculate its contribution to $Y'$ as follows:
\begin{equation}
S_W = IR_W \otimes S
\end{equation}
\begin{equation}
S_X = IR_X \otimes S \odot \cos(\theta) \sqrt{3}
\end{equation}
\begin{equation}
S_Y = IR_Y \otimes S \odot \sin(\theta) \sqrt{3}
\end{equation}

where $\otimes$ represents the convolution operation. To form $Y$, we treat $Y_L$ and $Y_R$ as two virtual sound sources positioned at azimuths of $\theta_L$ and $\theta_R$, respectively. With $Y$ at $\theta$ decoded using the polygon decoding method~\cite{zotter2019ambisonic}:

\begin{equation}
Y = Y_W + Y_X \cos(\theta) + Y_Y \sin(\theta)
\end{equation}

To satisfy $\frac{1}{2} (Y_L + Y_R) = Y_W$, $\theta_L$ and $\theta_R$ need to be $\pi$ radians apart. We enforce this during training and by setting $\theta_L = \frac{1}{8}\pi$ and setting $\theta_R = \theta_L + \pi = \frac{9}{8}\pi$. By synthesizing the Ambisonic data in this manner, we can generate a dataset of input-output pairs $(Y, Y')$ suitable for training the Ambisonizer model.

\subsection{Source Datasets}
\subsubsection{Ambisonic IR Datasets}
To generate synthetic spatial harmonics, we utilize Ambisonic IR datasets from well-established sources, namely OpenAIR~\cite{murphy2010openair}, Motus~\cite{gotz2021dataset}, and the C4DM RIR database~\cite{stewart2010database}. These datasets have been widely used in the research community for spatial audio applications and provide a diverse range of acoustic environments, including concert halls, studios, outdoor spaces, and indoor spaces with varying furniture layouts. To introduce variability and prevent overfitting to specific IR lengths, we randomly truncate each IR set to a duration between 0.3 and 1 second. This range is chosen based on the typical reverberation times encountered in real-world environments~\cite{kuttruff2016room}. By incorporating this randomization step, we ensure that our synthetic spatial harmonics are representative of a wide range of acoustic conditions that are realistic and may present in real-world recordings.

The truncation process is performed using a uniform random distribution to select the IR length within the specified range. This approach guarantees an unbiased sampling of IR durations while maintaining the integrity of the spatial information. Additionally, we apply a fadeout window to the truncated IRs to prevent abrupt endings and ensure a smooth transition. The fadeout length is randomly selected between 0.05 and 0.3 seconds.

\subsubsection{Sound Sources Datasets}

For the sound sources, we employ the MUSDB18-HQ dataset~\cite{rafii2017musdb18}, which consists of high-quality, multi-track recordings at 44.1 kHz of various musical genres. This dataset provides a diverse set of instruments and vocals, making it well-suited for generating artificial spatial mixes. To create these mixes, we randomly place each track in a virtual space by assigning azimuth values drawn from a uniform distribution between $-\pi$ and $\pi$ radians. For stereo tracks, we introduce a source width parameter, which is randomly selected from a range of 0 to $\pi$ radians. The left and right channels of the stereo track are placed symmetrically around the assigned azimuth, with the source width determining their angular separation.

To further enhance the variability and realism of the generated mixes, we apply a set of common audio augmentations to each track using the \texttt{audiomentations} library\footnote{\url{https://github.com/iver56/audiomentations}}. These augmentations include:

\begin{itemize}
    \item Gain adjustment: The gain of each track is randomly adjusted within a range of -10 dB to +10 dB with a probability of 70\%.
    \item Air absorption: A random air absorption effect is applied to simulate the frequency-dependent attenuation of sound over distance, with distances ranging from 0.1 to 10 meters and a probability of 70\%.
    \item Seven-band parametric equalizer: A seven-band parametric equalizer is applied to each track with random gain values between -12 dB and 12 dB for each band and a probability of 70\%.
    \item Gain transition: Smooth gain transitions in the range of -24 dB to 6 dB are introduced within each track with a duration between 0.2 and 6.0 seconds and a probability of 70\% to simulate dynamic volume automation in the mix.
\end{itemize}

\subsection{Experimental Setup}
We synthesized a training dataset consisting of 40 hours of audio at 44.1 kHz using the aforementioned synthesis pipeline. A random 10\% of this dataset was selected as a validation set for checkpoint selection during training. For both training and validation, we randomly cropped 120K samples (2.72 seconds) from each song, applied random gain, and fed the resulting data pairs into the model. The Ambisonizer model was trained for 800K steps with a batch size of 32, using the Adam optimizer with a cosine annealing schedule. The maximum and minimum learning rates were set to 5e-5 and 1e-7, respectively. We specified the latent representations $Z$ and $Z_C$ to have 64 dimensions. 

\section{Results}\label{sec:results}
After careful consideration, we have concluded that there is currently no established objective evaluation framework for our proposed Ambisonizer model. To the best of our knowledge, we were unable to find any existing Ambisonic upmixing baselines against which we could compare our model's performance, and there are no effective methods to objectively assess the \textit{plausibility} of generated Ambisonic recordings. Works in the Ambisonic field tend to not include objective evaluations as well~\cite{moreau20063d, berge2010new, barrett2012perception}, other than in the context of discussing decoding errors~\cite{murillo2014evaluation}.

One potential approach to evaluation could be to decode the generated Ambisonic audio to stereo and then apply the objective evaluation methods available in the stereo audio domain \cite{steinmetz2020auraloss}. However, this approach is not feasible for two reasons. Firstly, our proposed Ambisonizer model aims to generate a realistic Ambisonic sound field, which lacks the ability to encode creative yet unrealistic audio sources that are commonly found in popular stereo and surround recordings \cite{rumsey2012spatial, gibson2019art}. Secondly, introducing additional encoding and decoding stages that are not part of the Ambisonizer model itself would confound any metrics derived from the audio obtained after decoder rendering, making it difficult to isolate the model's performance \cite{bates2016comparing, narbutt2018ambiqual}.

Given the creative and subjective nature of our task, we believe that the most appropriate way to measure the model's performance is through subjective testing \cite{serra2023mono, bech2007perceptual, schoeffler2015towards}. To make the evaluation more accessible, we decode the Ambisonic recordings to stereo and compare our model's output against a strong commercial mono-to-stereo baseline. This approach allows for a more direct comparison of the perceived audio quality and spatial attributes, while still providing insight into the effectiveness of our Ambisonizer model in generating realistic Ambisonic signals. We detail our approach and findings below.

\subsection{Baseline: Waves PS-22}
\begin{figure}[hbt!]
    \centering
    \includegraphics[width=\linewidth]{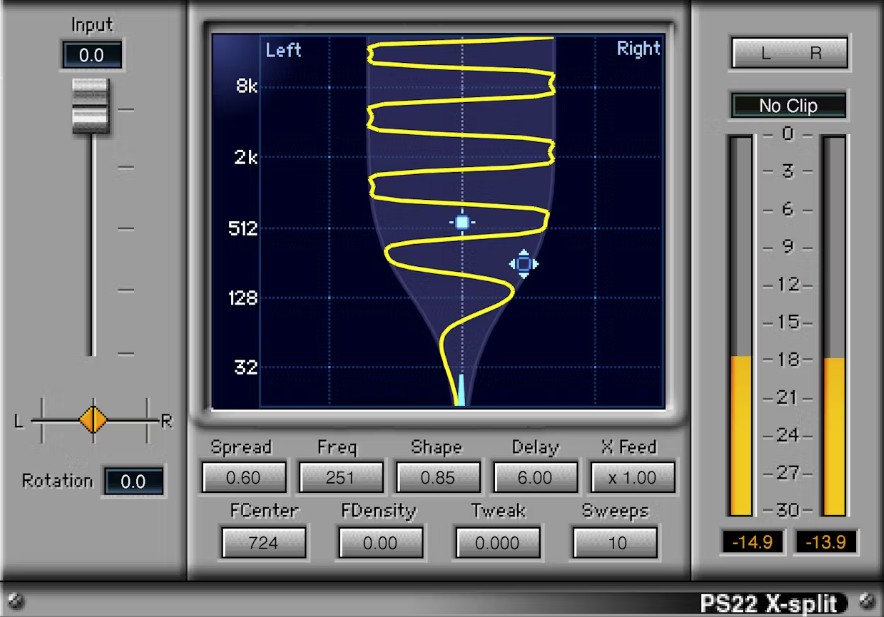}
    \caption{Waves PS-22 Stereo Maker\cite{waves2023ps22}}
    \label{fig:PS22}
\end{figure}

The demand for mono-to-stereo conversion is commercially significant. Introduced in 2012, the Waves PS-22 Stereo Maker (interface shown in Figure~\ref{fig:PS22}) has become a popular choice for this purpose. Many products rely on the Haas effect~\cite{liu2021identification}, which employs a short delay between the left and right signals to simulate a stereo field. However, this approach can lead to phase issues, resulting in cancellations between the left and right signals when they are downmixed to mono. In contrast, the PS-22 algorithm relies on a frequency distribution approach, assigning different frequencies to various stereo panning positions to achieve the stereo effect. This method avoids the phase cancellation issues associated with the Haas effect. As our Ambisonizer models require no human intervention, we employ the default settings of the Waves PS-22 in our experiments to ensure a consistent and reproducible approach to mono-to-stereo conversion.

\subsection{Subjective Evaluation}
\begin{figure*}
  \centering
  \includegraphics[width=\textwidth]{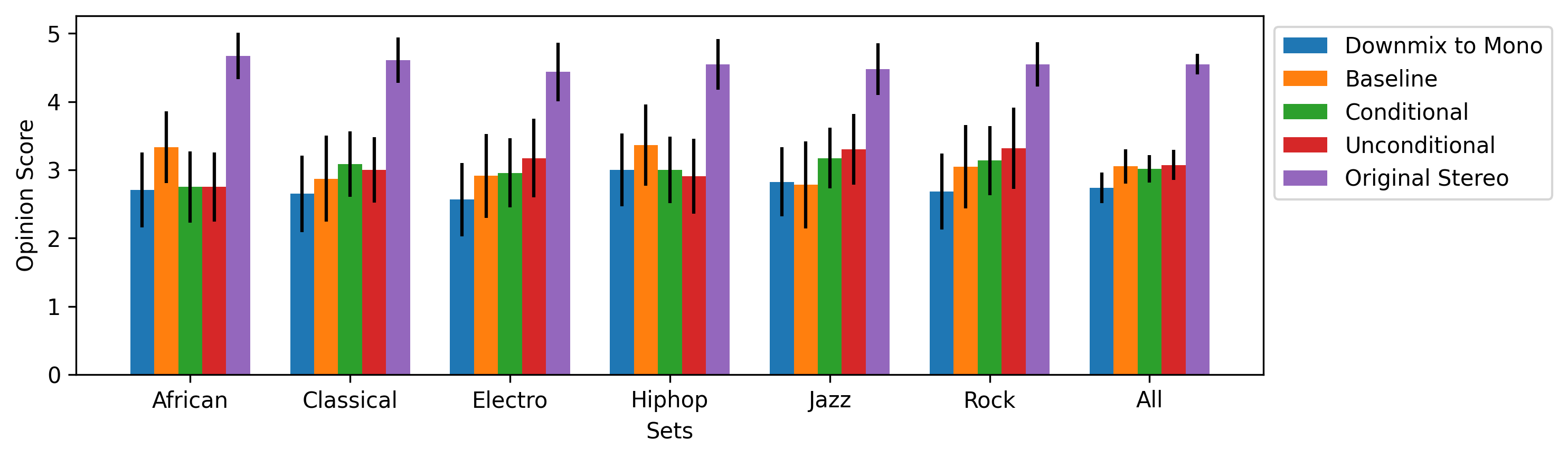}
  \caption{Subjective rating results. 'All' setting is calculated by aggregating all individual sets; error bars are calculated with a 95\% confidence interval.}
  \label{fig:ratings}
\end{figure*}

To form a comprehensive evaluation, we randomly selected 10-second audio samples from YouTube videos representing six genres: African\footnote{\url{https://www.youtube.com/watch?v=-1OmnxGo9pg}}, Classical\footnote{\url{https://www.youtube.com/watch?v=Sm4JaV6Xz0M}}, Electro\footnote{\url{https://www.youtube.com/watch?v=-GHtdr81VnE}}, Hiphop\footnote{\url{https://www.youtube.com/watch?v=u_wwfo4bs1o}}, Jazz\footnote{\url{https://www.youtube.com/watch?v=T75eWVt2OHI}} and Rock\footnote{\url{https://www.youtube.com/watch?v=syem-TmPTSo}}. While the MUSDB18 dataset prominently features the last four genres, the first two are not well-represented. We compared the performance of conditional and unconditional generation against the baseline, original stereo, and mono downmixed from stereo by collecting mean opinion scores (MOS) from participants. Each participant rated the five settings within each set, enabling direct comparison of the audio clips. To control for inherent fluctuations in subjective ratings, we followed the approach of~\cite{serra2023mono} and filtered out all set ratings where the mono downmix received a higher score than the original stereo recordings. The loudness of the audio clips was normalized to -24 LUFS to avoid loudness bias~\cite{vickers2010loudness}. For the Ambisonizer model settings, we conducted a grid search at 1-degree intervals for the left speaker position and 10-degree intervals for the distance between the left and right speakers. Same as during training, we used the regular polygon decoding method~\cite{zotter2019ambisonic} for a given speaker position. To account for the centered bias, as most stereo mixes have balanced left and right signals, we selected the two decoding speaker positions with minimal root-mean-squared (RMS) difference. 
A total of 25 participants provided 6 sets of ratings each. After applying the filtering rules, 13 sets were discarded from the analysis.

The results are illustrated in Figure~\ref{fig:ratings}. We additionally conduct a statistical significance test on all subjective ratings using pair-wise Wilcoxon signed-rank tests with $p = 0.05$. The results of the statistical significance test are shown in Table~\ref{tab: significance_test}.

\begin{table}[hbt!]
\centering
\begin{tabular}{c|cccc}
\hline\hline
              & Baseline & Cond. & Uncond. & Stereo \\ \hline
Mono          & Y        & Y           & Y             & Y      \\
Baseline      &          & N           & N             & Y      \\
Cond.   &          &             & N             & Y      \\
Uncond. &          &             &               & Y      \\ \hline\hline
\end{tabular}
\caption{Pairwise statistical significant test on all individual sets aggregated. N indicates no significant difference, while Y indicates a significant difference. Cond. and Uncond. stands for conditional and unconditional settings.}
\label{tab: significance_test}
\end{table}

Our analysis reveals that the baseline, conditional, and unconditional Ambisonizer models demonstrate statistically significant improvements over the mono downmix when considering the aggregated set. However, they perform worse compared to the stereo setting. No significant difference is observed among the baseline, conditional, and unconditional methods. Furthermore, we notice substantial fluctuations between each set. The baseline method shows strong performance on African and Hip-hop sets but underperforms compared to the Ambisonizer models on Classical, Electro, Jazz, and Rock settings.

\section{Discussions}

The results above demonstrate that the proposed Ambisonizer and its underlying paradigm are comparable to a strong commercial baseline. However, there remains a gap between the proposed approaches and the original stereo mixes. We hypothesize that this gap is due to three factors: 1) inherent limitations of the Ambisonic format, 2) decoding artifacts, and 3) the difficulty of the task itself. In this section, we present a detailed discussion of these aspects.

\textbf{Inherent limitations of the Ambisonic format.} As mentioned in Section~\ref{sec:results}, the Ambisonic format encodes audio in a realistic acoustic space and, therefore, cannot recreate more creative stereo and spatial effects. Upon listening to the produced renderings of the Ambisonizer model's output, we found that its perceived stereo field is narrower compared to the baseline and original stereo mix. We believe this issue may be addressed by developing additional decoder models, which take a reference input as a style guide during the decoding process.

\textbf{Decoding artifacts.} As Ambisonic decoding research progresses, we expect the decoding results to continue improving, thereby reducing the impact of decoding artifacts on the overall performance.

\textbf{Difficulty of the task itself.} The subjective nature of the task makes modeling what constitutes a \textit{plausible} upmixing result challenging. By upmixing to the Ambisonic format, we implicitly enforce that a \textit{plausible} result should, at the very least, be \textit{acoustically coherent}, meaning that all audio sources are in the same acoustic space. While some literature supports this notion~\cite{hepworth2016mixing}, it may not hold true for specific genres, such as electronic dance music (EDM).

Furthermore, we acknowledge two key limitations of our work. We posit that the worse performance on African and Hip-hop sets is likely due to the fact that these genres are percussion-heavy, and using a convolution-based audio decoder effectively sets a window, which makes modeling percussive attacks more difficult, as observed in source separation tasks~\cite{defossez2022high}. Additionally, we note that the lack of objective evaluation methods makes our work less meaningful due to the volatile nature of subjective ratings. By presenting our work and findings, we hope to inspire research into channel-agnostic upmixing paradigms, which would drive the development of objective evaluation methods and benchmarks. Despite these limitations, we believe the strong performance of the Ambisonizer model positions the Ambisonic format as a valid and promising intermediate representation for channel-agnostic upmixing.

\section{Conclusions}
We introduce Ambisonizer, a novel paradigm for channel-agnostic neural upmixing using spherical harmonics. By leveraging the Ambisonic format, our model enables the generation of first-order Ambisonic audio from mono or stereo input, allowing for mono-to-any and stereo-to-any upmixing. Through subjective evaluations, we demonstrated that the Ambisonizer model's output, when downmixed to stereo, is comparable to a strong commercial mono-to-stereo baseline. We also identified limitations in the Ambisonic format itself, decoding artifacts, and the inherent difficulty of the task. We believe that the strong performance of the Ambisonizer model positions the Ambisonic format as a promising intermediate representation for channel-agnostic upmixing.

\bibliography{ISMIRtemplate}
\end{document}